\documentstyle[prl,aps]{revtex}
\begin{document}
\draft
\twocolumn
\title{Distillability of Inseparable Quantum Systems}

\author{Micha\l{} Horodecki}

\address{Department of Mathematics and Physics\\
 University of Gda\'nsk, 80--952 Gda\'nsk, Poland}

\author{Pawe\l{} Horodecki}

\address{Faculty of Applied Physics and Mathematics\\
Technical University of Gda\'nsk, 80--952 Gda\'nsk, Poland}

\author{Ryszard Horodecki\cite{poczta}}

\address{Institute of Theoretical Physics and Astrophysics\\
University of Gda\'nsk, 80--952 Gda\'nsk, Poland}

\maketitle

\begin{abstract}
We apply the inseparability criterion for $2 \times 2$ systems,
local filtering and Bennett et al. purification protocol
[Phys. Rev. Lett. {\bf 76}, 722 (1996)]
to show how to distill {\it any}
inseparable $2\times 2$ system.
The extended protocol is illustrated geometrically 
by means of the state parameters in the Hilbert-Schmidt space.
\end{abstract}
%\newpage

Quantum error correction is one of the fundamental problems of the
quantum communication and quantum computation theory \cite{czw}. Within a
recently discovered  method of
transmission of quantum information 
(teleportation) \cite{BennettTel}, this can
be achieved indirectly by purification of an  ensemble of pairs of particles
used subsequently for asymptotically faithful teleportation \cite{Bennett}.
Namely,
Bennett et al (BBPSSW) \cite{Bennett} considered a protocol which allows
to obtain asymptotically a nonzero number of pairs of spin-$1\over2$ particles
in the singlet state from a large ensemble
described by a density matrix, provided that the latter has fidelity
greater than $1/2$. The fidelity is defined as \cite{czw}
\begin{equation}
f=\max \langle\psi|\varrho|\psi\rangle,
\end{equation}
where the maximum is taken over
all maximally entangled $\psi$'s.
The crux of the method is employment of
only local operations
and classical communication between Alice and Bob who share the particles
to be purified. Their protocol consists of
performing bilateral unitary transformations and measurements over
some number of pairs of particles.
(The obtaining singlet states
from ones under the two above conditions is called distillation or
purification).

A similar protocol was  used by Peres \cite{PeresBell} in collective tests for
nonlocality and by Deusch at al \cite{Deusch} in
the context of the security problem in quantum cryptography.

A way of obtaining more entangled states by using local operations and
classical communication has been proposed by Gisin \cite{Gisin}.
Similar method was used for concentrating of entanglement
for pure states by Bennett et al \cite{Bennett}. In Gisin's approach,
Alice and Bob subject the particles to the action of local filters,
and are able to obtain a mixture which violates Bell's inequality, despite
the fact that the original state satisfied them.

Note that the BBPSSW protocol cannot be applied to {\it all} inseparable
states. Indeed there are states with $f\leq 1/2$ which have nonzero
entanglement of formation \cite{czw} (hence cannot be written as convex
combinations of product states).
On the other hand, filtering method,
cannot be, in general, applied for direct production of singlets.
However, intuitively one feels that it should be possible to distill an
arbitrary inseparable state.
It involves a subtle problem of nonlocality of mixed states satisfying
standard Bell inequalities, first investigated
by Werner \cite{Werner} and Popescu \cite{Popescu1,Popescu2}.
Werner has found a family of states which are
inseparable (he called them EPR-correlated ones)
i. e. are not convex combinations of product states
but still admit the local hidden
variable model for single von Neumann measurement. He also conjectured that
the model exists also for POVM's. However Popescu showed that most of
Werner mixtures reveal nonlocality , if one takes into account the sequences
of measurements. Then he raised the question whether
all inseparable states are nonlocal.
This question could be solved just by showing that each inseparable state
can be distilled ( the distillability of the state shows it to be nonlocal).
The problem is that we do not have complete
``operational'' characterization of the inseparable mixed states.
Fortunately, quite recently, an effective criterion of inseparability
of mixed states for
$2\times2$ and $2\times3$ systems has been found \cite{PeresIns,sep}.

Here, using the criterion, filtering and BBPSSW
protocol
we will show that {\it any} inseparable mixed two spin-$1\over2$ state
can be purified to obtain assymptotically faithful teleportation.
In particular, as we shall see,
if one replaces filtering  by generalized measurements
(to avoid losing particles)  higher efficiency
of purification can be obtained by means a recursive process.

It has been shown \cite{PeresIns,sep}
that a state $\varrho$ of $2\times2$ system is inseparable
if and only if its partial transposition \cite{transp} is not a 
positive operator.
Suppose now that $\varrho$ is inseparable, and let $\psi$ be an eigenvector
associated with some negative eigenvalue of $\varrho^{T_2}$. Since in the 
process
of purification Alice and Bob can perform $U_1\otimes U_2$ transformations,
we can assume without loss of generality that
$\psi$ is of the form
\begin{equation}
\psi=a e_1\otimes e_1 +be_2\otimes e_2,
\end{equation}
where $\{e_i\}$ form the standard basis in $C^2$ and $a,b\geq0$.
Now $\langle\psi|\varrho^{T_2}|\psi\rangle<0$ implies
\begin{equation}
\langle I\otimes W\psi_2|\varrho^{T_2}|I\otimes W\psi_2\rangle<0
\label{lipa}
\end{equation}
where $\psi_2={1\over\sqrt2}(e_1\otimes e_1 +e_2\otimes e_2)$ and
\begin{equation}
W=
\left[\begin{array}{cc}
a&0\cr
0&b\cr
\end{array}\right] .
\end{equation}
Let us denote by $\tilde\varrho$ the state emerging after performing
the operation given by $I\otimes W$ 
\begin{equation}
\tilde\varrho={I\otimes W\varrho I \otimes W\over \text{Tr}
I\otimes W\varrho I \otimes W}.
\end{equation}
This state describes the subensemble of the pairs of particles,
which passed the local filter described by the operator $W$. Now
the inequality (\ref{lipa}) implies
\begin{equation}
\text{Tr}P_2^{T_2}\tilde\varrho<0,
\end{equation}
where $P_2=|\psi_2\rangle\langle\psi_2|$
(note that $P_2^{T_2}$ is equal to the operator $V$
given by $V\phi\otimes\tilde\phi=\tilde\phi\otimes\phi$, which was used by
Werner \cite{Werner} in the  necessary condition 
$\text{Tr}\varrho V\geq 0$ for separability).
However, the above inequality implies \cite{wyjasnienie}
\begin{equation}
\text{Tr}P_0\tilde\varrho>{1\over2},
\end{equation}
where $P_0$ denotes the singlet state  and
the state $\tilde\varrho$ can be purified by the BBPSSW protocol.

To summarize, given sufficiently many pairs of particles in an inseparable
state Alice and Bob can distill from it a nonzero number of singlets. To 
this end,
they first perform a measurement by means of complete set of product
observables on some number of particles, to get the matrix elements of
the state describing the ensemble
(it still involves only local operations and classical communication).
Then they perform a suitable of product unitary
transformations. Subsequently,
Alice directs her particles toward  a filter
the parameters of which can be derived from the density matrix
describing the ensemble.
Then Alice informs Bob, which particles have not been absorbed by  the filter,
so that he can discard the particles which lost their counterparts.
The subensemble obtained in this way can be now subjected to the
BBPSSW protocol to distill singlets. If the efficiency (the number of
purified pairs divided by the number of noisy pairs) of the latter
protocol is given by $\eta$, then the efficiency $\varepsilon$
of the whole process is given by
\begin{equation}
\varepsilon = \eta p,
\end{equation}
where $p=\text{Tr} (I\otimes W\varrho I\otimes W)$
is probability of passing the filter
i.e. the efficiency is product of the efficiencies of two stages:
filtering and BBPSSW protocol.

Although the  purification protocol described
above is effective in the sense that given any inseparable state  one can always
distill a nonzero number of singlets, it does not have to be the
best possible one.
It seems that for the inseparable states with $f\leq{1/2}$ the
best possible protocol should certainly consist of
filtering as the first stage, nevertheless, better efficiency of this
stage can be obtained. Consider for example the family of states
introduced
in the context of inseparability and Bell inequalities \cite{fazy}
\begin{equation}
\varrho=p|\psi_1\rangle\langle\psi_1| +
(1-p)|\psi_2\rangle\langle\psi_2|,
\label{st}
\end{equation}
where
$|\psi_1\rangle =ce_1\otimes e_1+de_2\otimes e_2  $,
$|\psi_2\rangle =ce_1\otimes e_2+de_2\otimes e_1$
where $c,d>0, \ p\not={1/2}$, and $\{e_i\}$ form the standard basis in $C^2$.
All the above states are inseparable.
Here, one should not follow the protocol described above, but rather
to apply the filter
\begin{equation}
W=
\left[\begin{array}{cc}
c&0\cr
0&d\cr
\end{array}\right].
\end{equation}

The efficiency of the first stage can be also raised by replacing the filter
with the generalized measurement one of the outcomes of which would produce
the same result as filtering.
The generalized measurement is given by a partition of unity
$\{V_i\}$, where $\sum V_i V_i^\dagger=I$.
After $i$-th outcome obtained (provided nondegeneracy of the measurement)
the state $\varrho$ collapses into
\begin{equation}
\varrho_i={ V_i \varrho V^\dagger _i\over
\text{Tr} (V_i \varrho V^\dagger_i)}.
\end{equation}
Thus instead of filter, one can use generalized measurement, and
choose the particles which produced suitable outcome $k$. The advantage here
is that if some other outcome was obtained, the particle is not lost
as in the case of filtering. It may be the case that the ensemble of the
particles which did not produce the required outcome would still be
described by some inseparable density matrix. Then one can
repeat the procedure, changing suitably the partition of unity,
to purify the subensemble. In this way
we obtain a recursive process, the efficiency
of which is higher than in the case of single filtering.

Now we will discuss our  purification protocol by means of
geometrical representation of the state  \cite{inf}.
For this purpose note that
an arbitrary two spin-$1\over2$ state can be represented in the
Hilbert-Schmidt (H-S) space
of all operators acting on $C^2\otimes C^2$
 as follows
\begin{equation}
\varrho={1\over4}(I\otimes I+\bbox {r\cdot\sigma}\otimes I+I\otimes\bbox
{s\cdot \sigma}+
\sum_{m,n=1}^3t_{nm}\sigma_{n}\otimes\sigma_{m}).
\label{postac}
\end{equation}
Here
$I$ stands for identity operator,
$ {\bbox r}$, ${\bbox s}$ belong to $R^3$,
$\{\sigma_n\}^3_{n=1}$ are the standard Pauli
matrices, $\bbox{ r\cdot\sigma}=\sum_{i=1}^3 r_{i}\sigma_{i} $.
The coefficients $t_{mn}={\rm Tr}(\rho\sigma_n \otimes \sigma_m)$
form a real matrix denoted by $T$.
The vectors $\bbox r$ and $\bbox s$ describes local properties of the state
while
the $T$ matrix describes a kind of projection of $\varrho$ onto
the set of states generated by maximally entangled projectors.
(see Ref. \cite{inf}
and references therein for more details concerning the formalism of the
H-S space of $2\times2$ system).
Thus the $T$ matrix determines whether the state can be directly
subjected to BBPSSW protocol to produce nonzero asymptotic singlets.
Indeed, basing on the results of Ref. \cite{inf} one obtains
 that  $f>1/2$ if and only if $N(\varrho)>1$ where
$N(\varrho)=\text{Tr}\sqrt{T^\dagger T}$, and then
\begin{equation}
f={1\over4}(1+N(\varrho)).
\end{equation}
For example, many of the states (\ref{st}) have $N(\varrho)\leq1$
hence they  {\it cannot} be purified by the BBPSSW protocol itself.
To find the Bell operator \cite{Mann} basis  in which a given state has the
highest fraction of
a maximally entangled vector, it suffices to find rotations which diagonalize
the $T$ matrix. Subsequently,  using the homomorphism between
the group unitary transformations of two level systems  and rotation group
 \cite{repr}, one can find the
suitable product unitary transformation which will convert the standard Bell
basis into the best one for the considered state.

Further,  we will assume that $T$ is diagonal so that
it can be treated  as a vector in $R^3$. It has been proven \cite{inf} that
if $\varrho$ is a state then $T$ must belong to the tetrahedron $\cal T$ with
vertices $(-1,-1,-1)$, $(-1,1,1)$, $(1,-1,1)$, $(1,1,-1)$
(see in this context \cite{czw}). Again, if $\varrho$ is
separable then $T$ must belong to the octahedron  $\cal L$ which is a
cross-section of $\cal T$ and $-\cal T$ (see fig. 1).

For the states with $\bbox r=\bbox s=0$ (we call them $T$-states)
the above conditions are also sufficient \cite{inf}, hence the set of $T$ states
is equal to the tetrahedron $\cal T$ and the set of separable $T$ states
can be identified with the octahedron $\cal L$ (note that $\cal L$
is described by inequality $N(\varrho)\leq 1 \cite{inf}$).

Consider now the following case, when the $T$ matrix of a given state lies
outside the octahedron (we will say that  the state lies outside the octahedron).
Then according to \cite{inf} there exists some maximally entangled
state $\psi$ such that $|\langle\psi|\varrho|\psi\rangle|>1/2$.
Thus, the state can be purified by the BBPSSW protocol.
Suppose now that the state lies inside the octahedron.
Then the first step of the BBPSSW protocol (random bilateral unitary
transformations) will destroy any inseparability of the state.
Indeed, there are two consequences of this step.
First one is that local parameters become $\bbox r=\bbox s=0$
(as a consequence of random rotations of vectors $\bbox r ,\bbox s$
inside of Bloch sphere). The second, very important one, is
that after the randomizing procedure
the $T$ matrix still remains inside the octahedron
(taking into account remarks from the previous paragraph  it is easy to see
that otherwise one could produce inseparable $T$-states from
separable $T$-states by use of local operations which
is obviously impossible).
Thus, according to the characterization of $T$-states,
the output state will be separable.

Now, the role of filtering becomes clear. Namely, this procedure
allows one to transfer the entanglement hidden in the relations between
$\bbox r$, $\bbox s$ and $T$ to the $T$ matrix itself.
If the input state is inseparable, but still lies inside
the octahedron, the process of filtering will move it outside it, so
that the BBPSSW protocol will produce a nonzero number of singlets.

Thus we have shown that any inseparable mixed two-spin-${1 \over 2}$
state can be distilled by using local operations
and exchange of classical information .
It solves completely the problem of nonlocality of $2\times2$ systems.

Finally, it is interesting to note that distillability
of an arbitrary inseparable mixed state of $2\times2$ system is 
exactly connected with the negative eigenvalue of partial transposition of
the state. Thus the possibility  of purification may be here interpreted as
a nonlocal effect ``produced'' by the eigenvalue.

\vskip1cm
We would like to thank Asher Peres for stimmulating discussions
and useful comments. We are also grateful Charles Bennett and
Sandu Popescu for useful remarks.


\begin{references}
\bibitem{transp}
The matrix elements of the partially transposed state are given by
$\varrho^{T_2}_{m\mu,n\nu}\equiv 
\langle e_m\otimes e_\mu| \varrho^{T_2}| e_n \otimes e_\nu\rangle=
\varrho_{m\nu,n\mu}$ where $\{e_i\}$ form the standard basis in $C^2$.
\bibitem{wyjasnienie}
This can be easily seen within the geometrical representation
of the state as discussed further.
\bibitem{repr}
The homomorphism is given by
$U\bbox{ \hat n\cdot\sigma}U^\dagger
=(O\bbox{\hat n})\bbox{\cdot\sigma}$ where
$U \in SU(2)$ and $O \in O^+(3)$ is a rotation.
Thus under product unitary transformation $U_1 \otimes U_2$
the parameters of the state transform as follows:
$\bbox {r}'=O_1 \bbox {r}$, $\bbox s'=O_2 \bbox s$, $T'=O_1 T O_2^\dagger$,
where $O_1$, $O_2$ are given by the homomorphism.


\begin{figure}
\caption{For the states with diagonal $T$ matrix the latter can be treated as
a vector in $R^3$. In particular, the projectors $\{P_i\}$  corresponding to
the Bell operator basis are uniquely represented by the points
 $A=(-1,-1,-1)$, $B=(1,1,-1)$, $C=(1,-1,1)$ and  $D=(-1,1,1)$.
Then i) for any state its  $T$ matrix must belong to
the tetrahedron $ABCD$ via the condition
$\text{Tr}\varrho P_i\geq0$; ii) for a separable state, $T$ must belong to
the bold-line-contoured octahedron, by virtue of the additional condition
$\text{Tr}\varrho P_i^{T_2}\geq0$.
Random bilateral unitary transformations ``project'' the $T$ matrix onto the
dashed line. For a state with $\text{Tr}\varrho P_0>1/2$ then the outputs of the
subsequent iterations of the BBPSSW protocol will lie on the line, closer
and closer to the singlet state represented by the point $A$.
\label{rys}}
\end{figure}
\end{references}
\end{document}